# Local pressure of confined fluids inside nanoslit pores
## (A density functional theory prediction)

by

F. Heidari, G. A. Mansoori[*] and E. Keshavarzi

*Department of BioEngineering, University of Illinois at Chicago, M/C 063, Chicago, IL 60607-7052, USA*

**Abstract**

In this work, the local pressure of fluids confined inside nanoslit pores is predicted within the framework of the density functional theory. The Euler-Lagrange equation in the density functional theory of statistical mechanics is used to obtain the force balance equation which leads to a general equation to predict the local normal component of the pressure tensor. Our approach yields a general equation for predicting the normal pressure of confined fluids and it satisfies the exact bulk thermodynamics equation when the pore width approaches infinity. As two basic examples, we report the solution of the general equation for hard-sphere (HS) and Lennard-Jones (LJ) fluids confined between two parallel-structureless hard walls. To do so, we use the modified fundamental measure theory (mFMT) to obtain the normal pressure for hard-sphere confined fluid and mFMT incorporated with the Rosenfeld perturbative DFT for the LJ fluid. Effects of different variables including pore width, bulk density and temperature on the behavior of normal pressure are studied and reported. Our predicted results show that in both HS and LJ cases the confined fluids normal pressure has an oscillatory behavior and the number of oscillations increases with bulk density and temperature. The oscillations also become broad and smooth with pore width at a constant temperature and bulk density. In comparison with the HS confined fluid, the values of normal pressure for the LJ confined fluid as well as its oscillations at all distances from the walls are less profound.

---

(*) Corresponding author: mansoori@uic.edu

# 1. Introduction

It is now well known that materials at a nanoscopic scale have sometimes interesting properties and behave quite different from than those in the macroscopic (bulk) scale. The major reason for the different behavior of confined fluids in nanoslit pores lies in the influence of system walls which significantly changes the energy and entropy of these systems as compared to macroscopic bulk fluid [1-8].

A particular trait of confined fluids in nanoslit pores is their inhomogeneous behavior which is revealed especially in density and consequently, affects all their properties. The pressure is an important property of confined fluids which is more complicated than the bulk pressure because of its tensorial character and its variations with respect to location and direction in a nanoslit pore [9-11]. Despite the importance of the knowledge about local pressure to study the behavior of nanoconfined fluids, there is a lack of appropriate experimental methods for pressure measurement and proper theoretical approaches for its prediction.

The major applications of the knowledge about local pressure in nanoconfined fluids are in derivation of the equations of motion of confined fluids, study of the interfacial properties, analysis of mechanical response to strain and heat, photo excitation and phase transformations [12].

There exist three approaches for calculation of pressure tensor: (i) The methods based on the hydrodynamics continuity equations such as the Irving-Kirkwood method [13] and the method of planes [9,14]; (ii) The virial theorem and integral equations such as the recent general equations derived for prediction of pressure tensor of confined fluids in nanoslit pores [15,6]; (iii) The approaches based on the density functional theory [16-18] that include the approach presented in this report.

In the present report we derive a general direct expression for predicting the local pressure of nanoconfined fluids within the framework of the density functional theory. Our approach is also applicable for the prediction of fluids pressure near interfaces and for fluids confined in micro-porous media. We use the modified fundamental measure theory (mFMT), which is shown to be one of the more successful versions of the DFT, to obtain the local pressure for hard-sphere confined fluid [19, 20]. Also, we use the mFMT incorporated within the Rosenfeld perturbative DFT [21] for long-range attractions, for local pressure prediction of the Lennard-Jones confined fluids in nanoslit pores.

This article is organized as follows: In Section 2, we present the derivation of the local normal component of the pressure tensor (called "normal pressure" for short). In Section 3 we report our solution to the local pressure theory for confined hard-sphere fluids in nanoslits. In Section 4 we report our solution for the local pressure of the Lennard-Jones nanoconfined fluid. Also reported in Section 4 are the effects of different variables, including temperature, pore width and bulk density, on local normal pressure for, both, hard-sphere and LJ fluids confined in nanoslits. Finally in Section 5 we present a discussion and our concluding remarks.

## 2. The Theory

The density functional theory (DFT) is a very successful and general approach for analyzing the behavior of inhomogeneous fluid systems in any size scale. While similar results may be obtained using molecular dynamics simulation, the DFT calculation is much faster than the MD simulation [4]. The density functional theory (DFT) is based on the idea that the Helmholtz free energy functional of an inhomogeneous fluid, $F$, and grand potential, $\Omega$, are expressed as functions of local density. The grand potential is related to its Helmholtz free energy functional via [22, 23]:

$$\Omega = F - \mu N, \tag{1}$$

where

$$F = F^{int} + \int \rho(\mathbf{r}) V_{ext}(\mathbf{r}) d\mathbf{r}, \qquad N \equiv \int \rho(\mathbf{r}) d\mathbf{r}, \tag{2}$$

and $\mu$ is chemical potential of the system, $V_{ext}(\mathbf{r})$ is the external field, $F^{int}$ is the intrinsic Helmholtz free energy functional and $\rho(\mathbf{r})$ is the equilibrium density distribution. According to the variational principle, the equilibrium density distribution function of a non-uniform fluid corresponds to the minimum of the grand potential with respect to density:

$$\left(\frac{\partial \Omega[\rho(\mathbf{r})]}{\partial \rho(\mathbf{r})}\right)_{T,\rho_b} = 0 \; ; \quad \left(\frac{\partial^2 \Omega[\rho(\mathbf{r})]}{\partial \rho^2(\mathbf{r})}\right)_{T,\rho_b} \geq 0, \tag{3}$$

which leads to the Euler-Lagrange equation,

$$\mu = \mu_{int}(\mathbf{r}) + V_{ext}(\mathbf{r}), \tag{4}$$

where $\rho_b$ is the bulk density in equilibrium with the confined fluid and

$$\mu_{int}(\mathbf{r}) = \left(\frac{\partial F^{int}[\rho(\mathbf{r})]}{\partial \rho(\mathbf{r})}\right)_{T,\rho_b}$$

is the intrinsic chemical potential.

Starting with the Euler-Lagrange equation for chemical potential is an approach for calculation of pressure tensor based on the force balance equation. The force equation describes the condition of static mechanical equilibrium in the system which may be obtained by taking the gradient from Eq. 4 as:

$$\nabla \mu_{int}(\mathbf{r}) + \nabla V_{ext}(\mathbf{r}) = 0, \tag{5}$$

where $\nabla$ symbolizes the spatial gradient operator. The right hand side of Eq. 5 is set equal to zero because chemical potential is constant and the spatial dependence of $\mu_{int}$ is exactly cancelled by the radial dependence of $V_{ext}(\mathbf{r})$. This is closely similar to the starting equation of Bartolotti and Parr's method to predict the local pressure in quantum mechanics and Percus's one in statistical mechanics. It has also been used in this article to predict the local pressure in nanoslit pore in a different approach.

Multiplying Eq. 5 by the local density yields the force density equation:

$$\rho(\mathbf{r})\nabla\left(\frac{\partial F^{int}[\rho(\mathbf{r})]}{\partial \rho(\mathbf{r})}\right)_{T,\rho_b} + \rho(\mathbf{r})\nabla V_{ext}(\mathbf{r}) = 0 \qquad (6)$$

The first term in the left-hand side of Eq. 6 is the sum of contributions of intermolecular interactions and kinetic energy of the system while the second term is contribution of the external potential. Integrating Eq. 6 over $\mathbf{r}$ will represent the total force exerted on a molecule at position $\mathbf{r}$ with equilibrium local density $\rho(\mathbf{r})$.

$$\int \rho(\mathbf{r})\nabla\left(\frac{\partial F^{int}[\rho(\mathbf{r})]}{\partial \rho(\mathbf{r})}\right)_{T,\rho_b} d\mathbf{r} + \int \rho(\mathbf{r})\nabla V_{ext}(\mathbf{r})d\mathbf{r} = 0 \qquad (7)$$

On the other hand, the force $F_j$ on a closed surface $s$ in direction $j$, is related to the stress tensor via [24]:

$$F_j = \oint_S \sigma_{ij}.n_i ds, \qquad (8)$$

where $\sigma_{ij}$ is the stress tensor and $n_i$ is the normal vector to surface. The closed surface integral in Eq. 8 may be transformed into a volume integral using the Gauss's theorem,

$$F_j = \oint_S \sigma_{ij}.n_i ds = \int_V \nabla_i \cdot \sigma_{ij} d\mathbf{r}. \qquad (9)$$

By comparing Eq. 9 with Eq. 7, we conclude:

$$\nabla_i \cdot \sigma_{ij} = -\rho(\mathbf{r})\nabla_j\left(\frac{\partial F^{int}[\rho(\mathbf{r})]}{\partial \rho(\mathbf{r})}\right)_{T,\rho_b} = -\rho(\mathbf{r})\nabla_j \mu_{int}(\mathbf{r}) = -\nabla_i \cdot \mathbf{P}_{ij}. \qquad (10)$$

Since Eq. 10 is an exact differential equation, we can rewrite it as:

$$d\mathbf{P}_{ii}(\mathbf{r}) = \rho(\mathbf{r})d\mu_{int}(\mathbf{r}). \qquad (11)$$

Eq, (11) may be integrated using integration-by-part as,

$$P_{ii}(\mathbf{r}) = \rho(\mathbf{r})\mu_{int}(\mathbf{r}) - \int \mu_{int}(\mathbf{r})d\rho(\mathbf{r}) + c(T, \rho_{bulk}), \qquad (12)$$

where $c(T, \rho_{bulk})$ is the constant of integration which depends on temperature and bulk density. Since the intrinsic chemical potential varies with position; $P_{ii}(\mathbf{r})$ is also an intrinsic local pressure. Considering that [22, 23]

$$F^{int}[\rho(r)] = \int f^{int}[\rho(r), \nabla\rho(r), r]dr, \quad \text{and} \quad \mu_{int}(\mathbf{r}) = \left(\frac{\partial F^{int}[\rho(\mathbf{r})]}{\partial \rho(\mathbf{r})}\right)_{T, \rho_b},$$

Eq. 12 is converted to:

$$P_{ii}(\mathbf{r}) = \rho(\mathbf{r})\mu_{int}(\mathbf{r}) - f^{int}[\rho(\mathbf{r})] + c(T, \rho_{bulk}), \qquad (13)$$

where $f^{int}$ is intrinsic Helmholtz free energy density of the system. By considering the macroscopic limit condition of Eq. 13 we conclude that the integration-constant $c(T, \rho_{bulk})$ will be equal to zero. Therefore, finally we have:

$$P_{ii}(\mathbf{r}) = \rho(\mathbf{r})\mu_{int}(\mathbf{r}) - f^{int}[\rho(\mathbf{r})]. \qquad (14)$$

Eq. 14 is a general equation for prediction of local pressure which is also applicable for confined fluids in nanopores. It should be noted that this equation also satisfies the macroscopic pressure when the size of confinement approaches infinity. In the macroscopic limit, the local densities are replaced with the average bulk density and intrinsic Helmholtz free energy density, $f^{int}$, is replaced with $\frac{F}{V}$. So, we get the following well-known general thermodynamic equation for macroscopic systems:

$$P = \frac{N\mu}{V} - \frac{F}{V} = \frac{G - F}{V}.$$

In the next section we use Eq. 14 to obtain the local pressure of hard-sphere confined fluid via the modified fundamental measure theory. In the subsequent section we extend our calculations to predict the local pressure of the Lennard-Jones nanoconfined fluid by incorporating the Rosenfeld perturbative density functional [25, 26] for long-range attractions.

## 3. Local normal component of the pressure tensor (normal pressure) of the hard-sphere fluid confined in a nanoslit pore

We consider here a nanoslit pore consisting of two structureless parallel hard (purely repulsive) walls in *xy* plane at *z*=0 and *z*=*H*, containing a hard-sphere (HS) fluid in equilibrium with a HS bulk fluid. Eq. 14 can then be used to obtain the local normal pressure of the confined HS fluid.

To derive expressions for $\mu_{int}(\mathbf{r})$ and $f^{int}(\rho(\mathbf{r}))$ which appear in the right-hand side of Eq.(14) we use the modified fundamental measure theory (mFMT) which is shown to be quite accurate [19]. In FMT formulation, the excess free energy contribution of the HS fluid is expressed in terms of fundamental geometrical measures of the particles [19-21]. In this model, the intrinsic Helmholtz free energy functional is expressed as the sum of ideal and excess parts, as:

$$F_{int}[\rho(\mathbf{r})] = F_{id}[\rho(\mathbf{r})] + F_{ex}[\rho(\mathbf{r})] \quad . \tag{15}$$

The ideal-gas contribution of the Helmholtz free energy is [22]:

$$F_{id}[\rho(\mathbf{r})] = k_B T \int \rho(\mathbf{r}) \{\ln \rho(\mathbf{r}) \Lambda^3 - 1\} d\mathbf{r} \quad , \tag{16}$$

where $k_B$ is the Boltzmann constant, $T$ is the absolute temperature and $\Lambda$ is the thermal de Broglie wavelength. The excess part of the Helmholtz free energy, $F_{ex}[\rho(\mathbf{r})]$, takes into account the non-ideality due to intermolecular interactions and we obtain it using the modified fundamental measure theory (mFMT) proposed by Yu and Wu [19]. It is based on the accurate MCSL equation of state for pure and mixture hard-sphere fluids [27]. According to mFMT, the excess part of the Helmholtz free energy is:

$$F_{ex}[\rho(\mathbf{r})] = k_B T \int \left( \Phi_S^{hs}[n_\alpha(\mathbf{r})] + \Phi_V^{hs}[n_\alpha(\mathbf{r})] \right) d\mathbf{r}, \tag{17}$$

where $\Phi^{hs}[n_\alpha(\mathbf{r})]$ is the excess Helmholtz free energy density of a HS fluid which is a function of the weighted density distribution, $n_\alpha(\mathbf{r})$. Subscripts $S$ and $V$ in Eq. 17 indicate scalar and vector parts, respectively. According to mFMT, the two parts of $\Phi^{hs} = \Phi_S^{hs} + \Phi_V^{hs}$ are given as [19-21]

$$\Phi_S[n_\alpha(\mathbf{r})] = -n_0 \ln(1-n_3) + \frac{n_1 n_2}{1-n_3} + \left[ \frac{1}{36\pi n_3^2} \ln(1-n_3) + \frac{1}{36\pi n_3 (1-n_3)^2} \right] n_2^3 \quad , \tag{18}$$

and

$$\Phi_V[n_\alpha(\mathbf{r})] = -\frac{\mathbf{n}_{v1} \cdot \mathbf{n}_{v2}}{1-n_3} - \left[ \frac{1}{12\pi n_3^2} \ln(1-n_3) + \frac{1}{12\pi n_3 (1-n_3)^2} \right] n_2 (\mathbf{n}_{v2} \cdot \mathbf{n}_{v2}) \quad , \tag{19}$$

and the weighted density distribution is defined as:

$$n_\alpha(\mathbf{r}) = \int d\mathbf{r}' \rho(\mathbf{r}') w^{(\alpha)}(\mathbf{r} - \mathbf{r}') \quad , \tag{20}$$

where $\alpha = 0, 1, 2, 3, v_1$ and $v_2$. The first four weights are scalar and the last two are vector weight functions.

For the HS fluid in nanoslit pore confined with two parallel-structureless hard walls, the external potential and density profile vary only in the z-direction (perpendicular distance from the wall), i.e. $\rho(\mathbf{r}) = \rho(z)$, $V_{ext}(\mathbf{r}) = V_{ext}(z)$. The external purely repulsive potential is then has the following form:

$$V_{ext}(\mathbf{r}) = V_{ext}(z) = \begin{cases} 0 & \text{for } \frac{\sigma}{2} < z < H - \frac{\sigma}{2} \\ \infty & \text{for } H - \frac{\sigma}{2} \leq z \leq \frac{\sigma}{2} \end{cases} \quad (21)$$

where $H$ is the width of the nanoslit and $z$ is the perpendicular distance from the wall (see Figure 1).

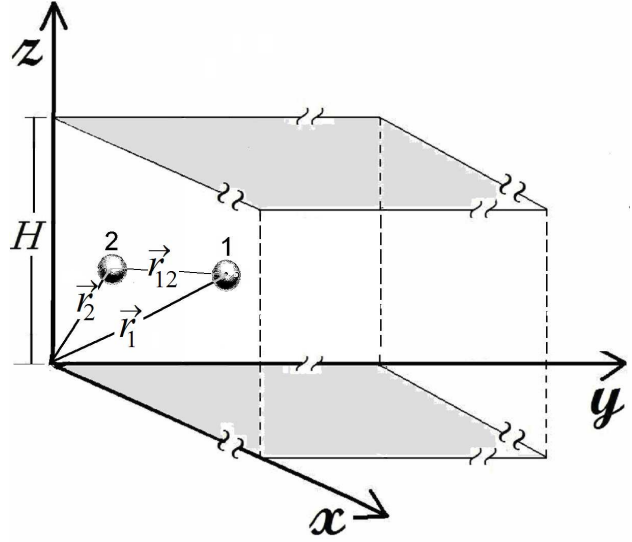

**Figure 1.** Illustration of a nanoslit consisting of two structureless parallel walls located at $z=0$ and $z=H$ and two confined particles, 1 and 2.

Therefore the four weighted scalar density distribution are:

$$n_0(z) = \frac{2n_1(z)}{\sigma}, \quad (22)$$

$$n_1(z) = \frac{n_2(z)}{2\pi\sigma}, \quad (23)$$

$$n_2(z) = \pi\sigma \int_{-\sigma/2}^{+\sigma/2} \rho(z+z')dz', \quad (24)$$

$$n_3(z) = \pi \int_{-\sigma/2}^{+\sigma/2} \rho(z+z')\left[\frac{\sigma^2}{2} - z'^2\right]dz', \quad (25)$$

and the two vector weighted density distribution are

$$\mathbf{n}_1(z) = \frac{\mathbf{n}_2(z)}{2\pi\sigma}, \tag{26}$$

$$\mathbf{n}_2(z) = 2\pi\mathbf{e}_z \int_{-\sigma/2}^{+\sigma/2} \rho(z+z')z'dz', \tag{27}$$

and Eq. 14 is simplified to:

$$P_{zz}(z) = \rho(z)\mu_{int}(z) - f^{int}(\rho(z)). \tag{28}$$

To solve Eq. 28, at first we need the equilibrium density distribution of nanoslit pore, $\rho(z)$, based on which the intrinsic Helmholtz free energy density and intrinsic chemical potential may be obtained. The equilibrium density distribution in nanoslit geometry is given by Eq. 29, which is obtained by minimizing the grand potential and it is calculated using the Picard iterative method [4]:

$$\rho(z) = \rho_b \exp\left(\beta\mu_{ex}^{HS} - \int dz' \sum_\alpha \frac{\partial \Phi^{HS}}{\partial n^\alpha(z)} w^{(\alpha)}(z-z') - \beta V_{ext}(z)\right), \tag{29}$$

where $\beta \equiv (k_b T)^{-1}$ and $\mu_{ex}^{HS}$ is the excess hard-sphere chemical potential (over ideal gas) obtained using the Carnahan-Starling equation of state [28] which is the same as MCSL equation of state [27] for pure hard-sphere fluid:

$$\beta\mu_{ex}^{HS} = \frac{\eta(8 - 9\eta + 3\eta^2)}{(1-\eta)^3}, \tag{30}$$

and where $\eta = (\pi/6)\rho\sigma^3$ is the packing fraction of the fluid.

The Helmholtz free energy density, Eq. 17, can be calculated by replacing Eq.s 18 and 19 in Eq. 17. The intrinsic Helmholtz free energy derivative with respect to density also yields the intrinsic chemical potential according to Eq. (12-2). Consequently, the local normal pressure of the hard-sphere fluid confined in a nanoslit pore can be calculated via Eq. 28.

To test the accuracy of our results, we have calculated the wall pressure for hard-spheres confined in nanoslit pores and we have made comparisons between wall pressures resulting from present approach with the Monte Carlo simulations [29] at the reduced bulk density $\rho^* = \rho\sigma^3 = 0.7$ as it is shown in Figure 2.

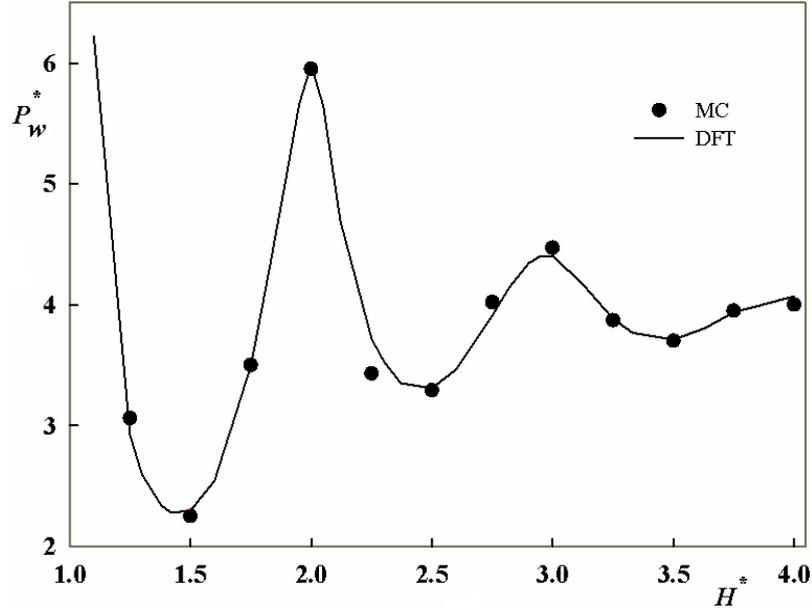

**Figure 2.** The wall pressure ($P_W^* = P_W \cdot \sigma^3 / kT = \rho_{H=0}\sigma^3 = \rho_{H=0}^*$) as a function of dimensionless pore width ($H^* = H/\sigma$) for hard-spheres confined in slit pores at the reduced bulk density $\rho_b^* = \rho_b \sigma^3 = 0.7$. The points are from Monte Carlo simulation [30], the solid line is our predictions.

According to the contact-value theorem, the contact density in a nanoslit pore is directly proportional to the wall pressure $P_w^*$.

We have also made a comparison between the density distribution at the flat wall obtained via DFT and the Monte Carlo simulation results [30, 31] for a hard-sphere fluid confined in a nanoslit pore at the reduced density of $\rho^* = 0.8$ as it is shown in Figure 3.

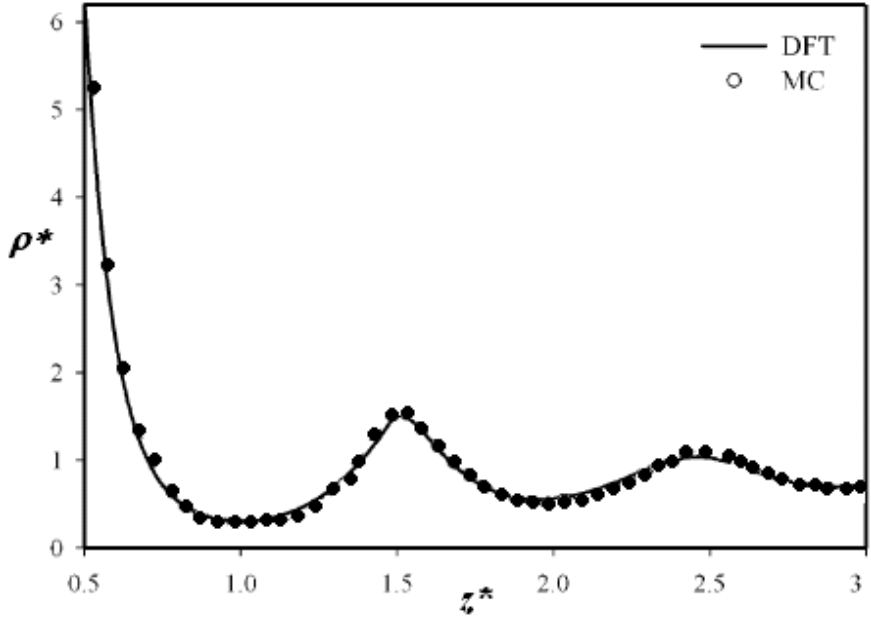

**Figure 3.** Comparing the reduced density distribution at the flat wall obtained via DFT and the Monte Carlo simulation results [31] for a hard-sphere fluid with the reduced bulk density of $\rho_b^* = \rho_b \sigma^3 = 0.8$ as a function of dimensionless distance from the wall $z^* = z/\sigma$

It is clear from Figures 2 and 3 that the agreement between our DFT calculation results and the MC simulation data are quite good. It should be noted that the density profile in the nanoslit pore has an oscillatory behavior as shown in Figure 3. Density oscillatory behavior indicates fluid layering structure in nanoslits. Density oscillations influence all the structural properties of confined fluids including the oscillatory behavior of normal pressure as it is shown below.

In Figure 4 we report our DFT calculation for the local normal pressure of the hard-sphere fluid confined in the nanoslit pore at the reduced temperature $T^* = kT/\varepsilon = 2.5$, reduced bulk density $\rho_b^* = \rho_b \sigma^3 = 0.6$ and for two different dimensionless pore widths $H^* = H/\sigma = 4$ and 6.

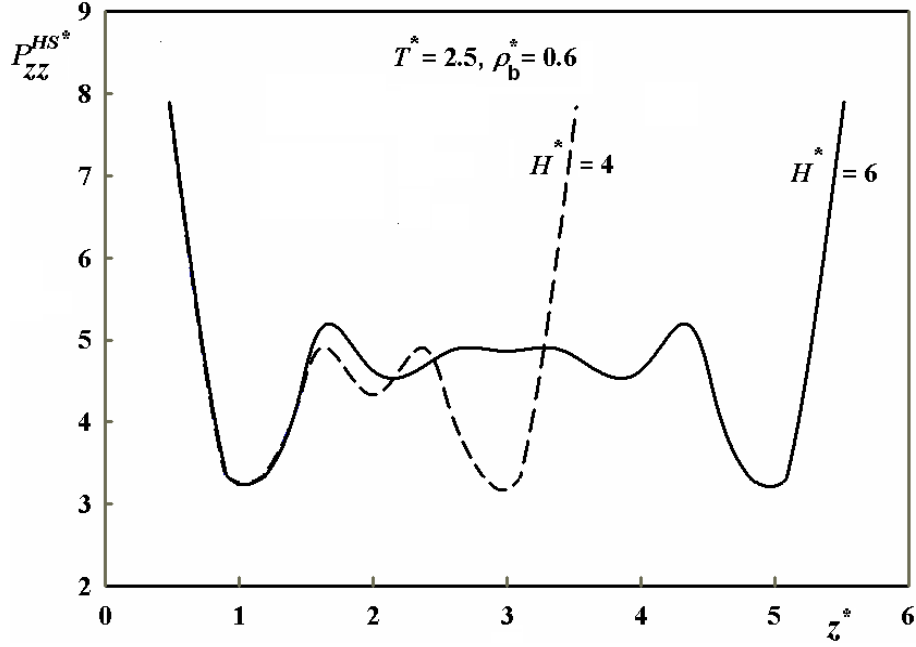

**Figure 4.** The reduced normal component of the pressure tensor (normal pressure) $P_{zz}^{HS*} = P_{zz}^{HS}\sigma^3/\varepsilon$ for hard-sphere confined fluids in nanoslit pore versus dimensionless distance from the wall, $z^* = z/\sigma$, made of two parallel-structureless hard walls at $\rho_b^* = \rho_b\sigma^3 = 0.6$ and $T^* = kT/\varepsilon = 2.5$ for two different reduced pore widths of $H^* = H/\sigma = 4$ & 6.

In defining the reduced pressure and reduced temperature we use $\varepsilon = 119.8k$ which is the Lennard-Jones energy parameter value for argon-argon interaction and $k$ is the Boltzmann constant. According to Figure 4 the local normal pressure of the hard-sphere confined fluid has a symmetrical oscillatory behavior versus $z^*$ which decreases from the walls. According to Figure 4 for a nanoslit with $H^* = 4$ and $\rho_b^* = \rho_b\sigma^3 = 0.6$ the hard-sphere fluid forms two layers. As the slit width increases the number of layers also increases. For $H^* = 6$ the hard-sphere molecules have enough space to form six distinct layers, two sharp ones at the walls and four broader layers in the middle.

In Figure 5 we report the local normal pressure of the hard-sphere confined fluid at $T^* = 2.5$ and $H^* = 6$ for two different reduced bulk densities $\rho_b^* = \rho_b\sigma^3 = 0.6$ and 0.8.

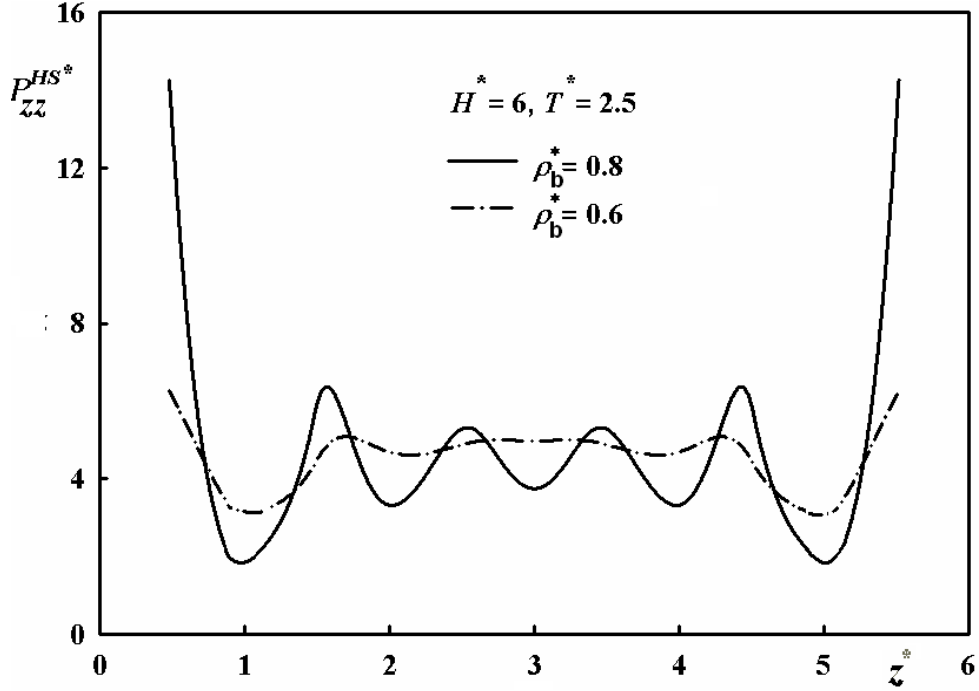

**Figure 5.** The reduced normal pressure for hard-sphere confined fluids in nanoslit pores $P_{zz}^{HS*} = P_{zz}^{HS}\sigma^3/\varepsilon$ with hard walls versus dimensionless distance from the wall, $z^* = z/\sigma$, at $H^* = H/\sigma = 6$ and $T^* = kT/\varepsilon = 2.5$ for two different bulk reduced bulk densities of $\rho_b^* = \rho_b\sigma^3 = 0.6$ & $0.8$.

According to this figure, at constant temperature and pore width when the bulk density, and as a result the average density in nanoslit pore, increase, molecules show a stronger tendency to accumulate at the walls. Therefore, the normal pressure as well as the height and depth of oscillation increase at distances closer to the walls.

To investigate the behavior of local normal pressure of a realistic model for simple molecular fluids, we have also studied the Lennard-Jones (LJ) fluid as described below.

## 4. Local normal component of the pressure tensor (normal pressure) of the confined Lennard-Jones (LJ) fluid in nanoslit pores

To investigate the effect of long-range intermolecular attractions on the normal pressure, we consider the LJ confined fluid in nanoslit pores with two structureless parallel hard (purely repulsive) walls. We assume that the system consists of $N$ molecules confined between two parallel hard walls with macroscopic areas and nanoscopic pore width equal to $H$.

According to the perturbative density functional theory [25, 26], the excess part of intrinsic Helmholtz free energy of the LJ fluid includes both repulsive and attractive contributions. We have incorporated the excess Helmholtz free energy functional in terms of mFMT [19, 20] for hard-core repulsion and the mean-spherical-approximation for the remaining long-range intermolecular attractive energy contribution. According to this approach, we have:

$$F_{ex}[\rho(\mathbf{r})] = F_{ex}^{hc}[\rho(\mathbf{r})] + F_{ex}^{att}[\rho(\mathbf{r})], \tag{31}$$

where the superscripts 'hc' and 'att' represent the hard-core reference and attractive contributions to the Helmholtz free energy functional, respectively. The excess Helmholtz free energy functional, due to the long-range attraction, can be expressed using the analytical expression of the direct correlation function from the first order mean spherical approximation [32, 33].

$$F_{ex}^{att}[\rho(\mathbf{r})] = -\frac{kT}{2} \iint \rho(\mathbf{r}_1)\rho(\mathbf{r}_2) c^{att}(|\mathbf{r}_2 - \mathbf{r}_1|) d\mathbf{r}_1 d\mathbf{r}_2, \tag{32}$$

where $c^{att}(|\mathbf{r}_2 - \mathbf{r}_1|)$ is the attractive part of the direct correlation function, which can be expressed as [33]

$$c^{att}(\mathbf{r}) = \begin{cases} c^Y(T_1^*, z_1 d, r/d) - c^Y(T_2^*, z_2 d, r/d) &, \quad r < d \\ 0 & \quad d \leq r < \sigma \\ -\frac{4}{T^*}\left(\left(\frac{\sigma}{r}\right)^{12} - \left(\frac{\sigma}{r}\right)^6\right) & \quad \sigma < r < r_c \end{cases} \tag{33}$$

The first order Yukawa direct correlation function, $c^Y$, in Eq. 33 is as follows:

$$c^{att}(\mathbf{r}) = \frac{1}{T_1^*}\left[\frac{\exp(-t_1(r-\sigma)/\sigma)}{r/\sigma} - Q(t_1)P(r,t_1)\right] - \frac{1}{T_2^*}\left[\frac{\exp(-t_2(r-1)/\sigma)}{r/\sigma} - Q(t_2)P(r,t_2)\right], \tag{34}$$

with a set of functions which are described as follow:

$$Q(t) = [S(t) + 12\eta L(t)e^{-t}]^{-2} \tag{35}$$

$$S(t) = \Delta^2 t^3 + 6\eta\Delta t^2 + 18\eta^2 t - 12\eta(1+2\eta) \tag{36}$$

$$L(t) = (1+\eta/2)t + 1 + 2\eta \tag{37}$$

$$\begin{aligned}P(r,t) &= S^2(t)\frac{e^{-t(r-\sigma)/\sigma}}{r/\sigma} + 144\eta^2 L^2(t)\frac{e^{-t(r-\sigma)/\sigma}}{r/\sigma} - 12\eta^2[(1+2\eta)^2 t^4 + \Delta(1+2\eta)t^5]r^3/\sigma^3 \\ &+ 12\eta[S(t)L(t)t^2 - \Delta^2(1+\eta/2)t^6]r/\sigma - 24\eta[(1+2\eta)^2 t^4 - \Delta(1+2\eta)t^5] \\ &+ 24\eta S(t)L(t)\sigma/r\end{aligned} \tag{38}$$

where $\Delta = 1-\eta$ and $\eta = (\pi/6)\rho d^3$ is the packing fraction. By using Eq. 32, the attractive part of the excess Helmholtz free energy will be obtained. Its combination with the hard-core reference system will generate the intrinsic Helmholtz free energy of the LJ fluid. Then, by having the equilibrium density distribution of the LJ confined fluid which is calculated using an iterative method using the following equation; its normal pressure will be obtained using Eq. 28.

$$\rho(z) = \rho_b \exp\left(\beta\mu_{ex} - \int dz' \sum_\alpha \frac{\partial \Phi^{HS}}{\partial n^\alpha(z)} w^{(\alpha)}(z-z') + \int dz' \rho(z') c^{att}(z) - \beta V_{ext}(z)\right).$$

The behavior of local normal pressure of the LJ confined fluid as calculated through the above mentioned method is reported in Figures 6- 9. In Figure 6 we report the results of our calculation for the normal pressure profiles of the LJ confined fluid in nanoslit pores at $T^* = 2$ and bulk density $\rho^* = 0.6$ for two different pore widths $H^* =$ **4** and **6.**

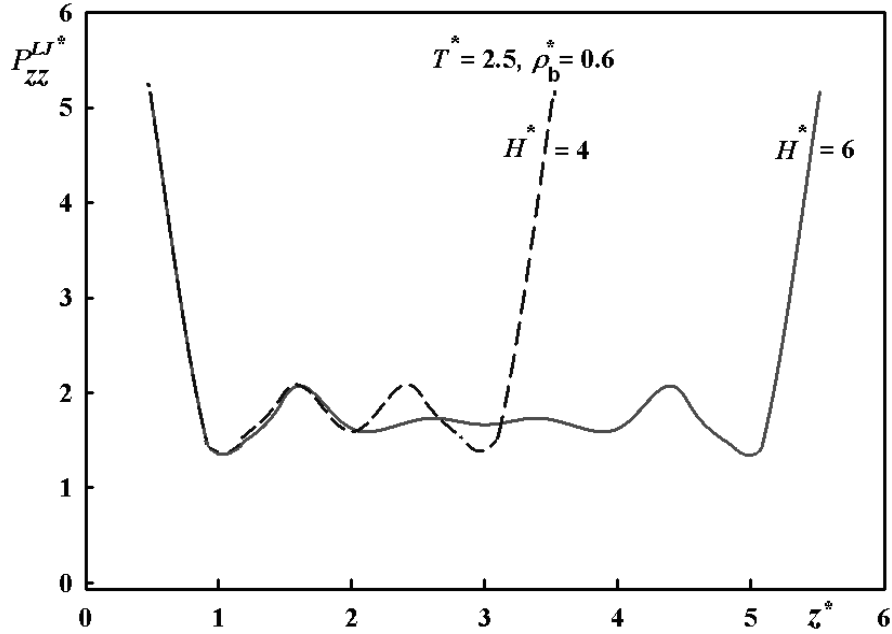

**Figure 6.** The reduced normal component of the pressure tensor (normal pressure) $P_{zz}^{LJ*} = P_{zz}^{LJ}\sigma^3/\varepsilon$ for the Lennard-Jones confined fluid in nanoslit pore versus dimensionless distance from the wall, $z^* = z/\sigma$, made of two parallel-structureless hard walls at $\rho_b^* = \rho_b\sigma^3 = 0.6$ and $T^* = kT/\varepsilon = 2.5$ for two different reduced pore widths of $H^* = H/\sigma = 4$ & 6.

According to this figure, the increase in pore width leads to an increase in the number of layers similar to the hard-sphere case reported in Figure 4. However, the LJ pressures are much smaller and less sensitive to slit width than the corresponding HS pressures due to the attractive energy btween the Lennard-Jones molecules.

In Figure 7 we report the local normal pressure of the LJ confined fluid at $T^*=2$ and 2.5 and $H^*=6$ for two different bulk densities $\rho_b^*=0.6$ and 0.8.

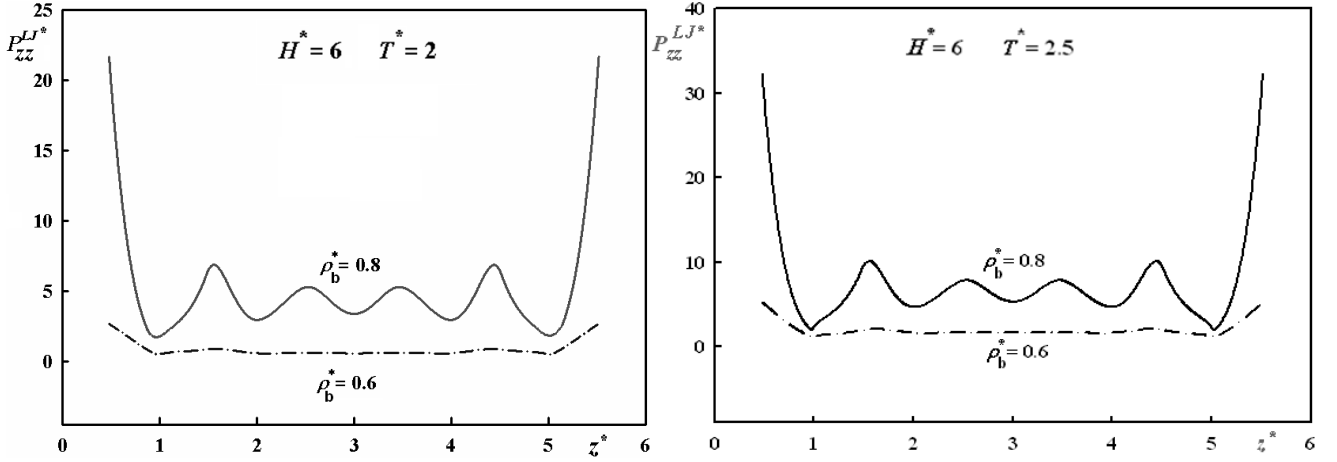

**Figure 7.** The reduced normal component of the pressure tensor (normal pressure) $P_{zz}^{LJ*} = P_{zz}^{LJ}\sigma^3/\varepsilon$ for the Lennard-Jones confined fluid in nanoslit pore versus dimensionless distance from the wall, $z^* = z/\sigma$, made of two parallel-structureless hard walls at $H^* = H/\sigma = 6$ and two different reduced temperatures $T^* = kT/\varepsilon = 2$ and 2.5 and two different reduced densities $\rho_b^* = \rho_b\sigma^3 = 0.6$ and 0.8.

According to Figure 7 when the bulk density increases, the local normal pressure of the LJ fluid confined increases. The increase in density (at constant temperature and pore width) leads to an increase in the number of layers similar to the hard-sphere case reported in Figure 5. Also according to Figure 7, the LJ pressure at $\rho_b^* = \rho_b\sigma^3 = 0.8$ are larger than pressure at $\rho_b^* = 0.6$ as compared with the corresponding HS case in Figure 5. This is an indication that there is a stronger tendency for LJ molecules than HS molecules to enter the nanoslit when bulk density increases.

Also, Figures 6 and 7 show the high sensitivity of the oscillatory behavior of normal pressure of the LJ fluid with increases in bulk density and pore width at constant temperature.

The data for $\rho_b^* = \rho_b\sigma^3 = 0.6$ of Figure 7 is plotted with higher resolution for the vertical coordinate in Figure 8 to demonstrate the effect of temperature on the normal pressure component of the pressure tensor in a nanoslit pore.

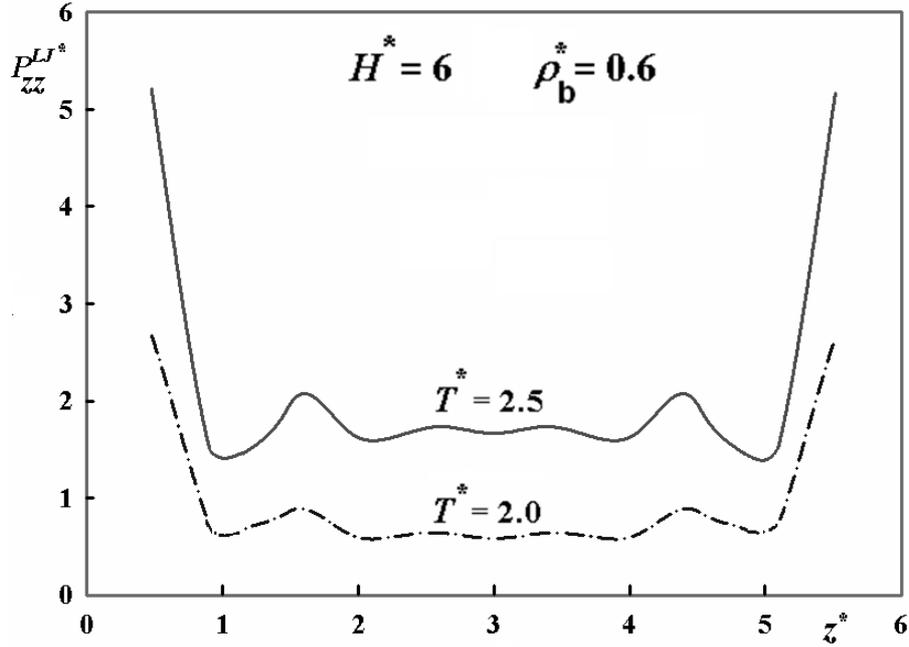

**Figure 8.** The normal pressure for LJ confined fluids in nanoslit pore with hard walls at $\rho_b^* = 0.6$ and $H^* = H/\sigma = 6$ for two different reduced temperatures $T^* = kT/\varepsilon = 2$ and 2.5.

According to Figure 8 at $\rho_b^* = \rho_b \sigma^3 = 0.6$ and $H^* = H/\sigma = 6$, the tendency for molecules to accumulate at the walls have been increased with increasing temperature or decreasing intermolecular attraction. Also, according to Figure 8 the values of local pressure increase and its oscillations also become more profound as temperature increases.

In order to demonstrate the differences between the normal pressure of confined hard-sphere and Lenard-Jones we report figure 9.

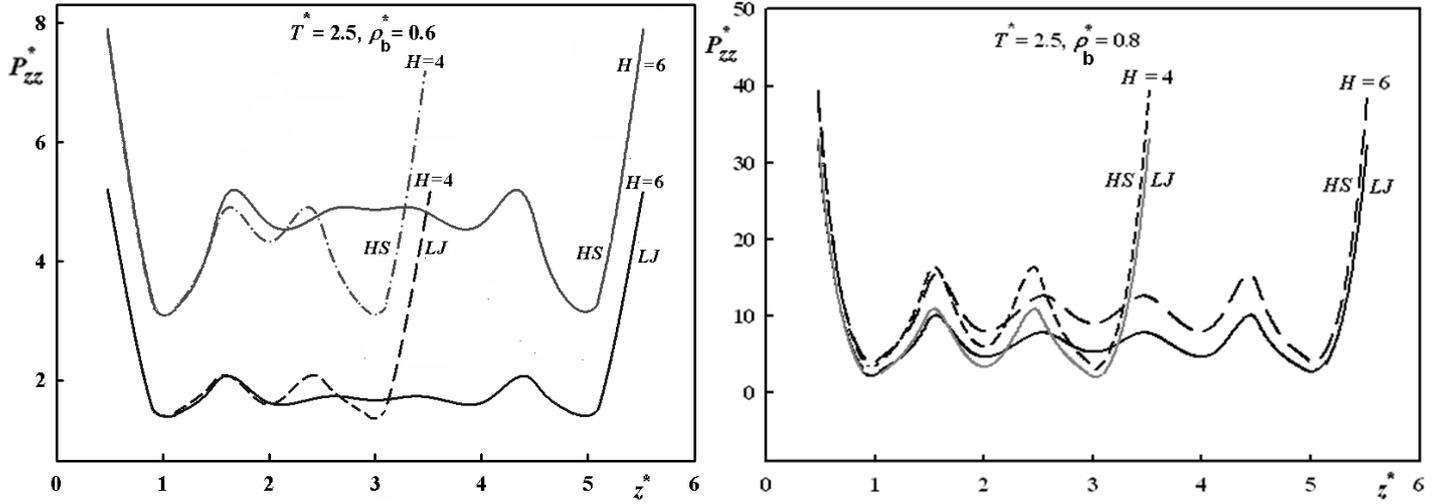

**Figure 9.** Comparison of local pressure of LJ confined fluid with those obtained for hard-sphere at $\rho_b^* =0.6$ and 0.8 and $T^* = 2.5$ for two different pore widths equal to $H^*=4$ and 6.

According to Figure 9 the normal pressure of both cases have oscillatory behavior versus distances from the walls. However, the values of normal pressure for LJ confined fluid as well as its oscillations at all distances from the walls are less profound than the hard-sphere ones because of the presence of attractive forces in the Lennard-Jones.

## 5. Results and discussion

As we have shown in this paper, the presented method is a general equation based on the density functional theory (DFT) which can be applied to predict the local normal pressure for any confined fluid with different geometrical walls. The presented equation has been directly extracted from Euler-Lagrange equation in density functional theory?. As shown in this article, DFT is the convenient way to predict the local behavior of confined fluid, while in the other approach some approximation has been used which is not applicable at high density.

As a result, the normal pressure of LJ confined fluid is the same as hard-sphere except that the value of normal pressure profiles of LJ fluid for all values of $z^*$ as well as its oscillatory behaviors are much less profound than the normal pressure profiles of hard-sphere.

This indicates that while the repulsive intermolecular interaction determines the general behavior of the normal pressure, its attractive part plays a significant role in value of the normal pressure of confined fluid. The long-range attractive intermolecular interactions hold back the molecules due to cutting off of the interaction by the walls which have a more important role at lower temperature.

It would also be interesting to compare the presented equation with the recent equation proposed by Keshavarzi *et al.* in Ref.s [6] and [15]. Although using two different definitions for pressure tensor is the main difference between these two methods leading to the different numerical results, both of these methods predict the oscillatory behavior of normal pressure and the effect of pore width, bulk density and temperature on its behavior is similar. Both methods are applicable for any fluid that is confined in a nanoslit pore and with any kind of intermolecular interactions. They also satisfy the macroscopic thermodynamic equation when the pore width approaches infinity.

## 6. Conclusion

In this work, we have derived a general expression for predicting the normal component of the pressure tensor (normal pressure) of confined fluid within the framework of the density functional theory of statistical mechanics. This approach is a direct consequence of the Euler-Lagrange equation which leads to a simple and efficient equation to calculate the normal pressure of confined fluid in nanoslit pore. It can also be generalized to calculate the local pressure in other geometries of nanopore.

Our results show that the normal pressure of a fluid in nanopore is a function of the pore width, intermolecular interaction and wall interaction as well as temperature and bulk density. Also, the normal pressures of both examined confined fluids (hard-sphere and Lennard-Jones) possess oscillatory behavior versus distance from the walls which increase with temperature and bulk density. Increasing temperature causes an increase in the tendency of molecules to enter the nanoslit. In addition, in the case of the LJ fluid, increasing temperature may be considered as reducing the well depth of potential, so tendency of molecules increases to accumulate at the walls. It has been observed that at constant temperature and pore width, the oscillatory behavior of normal pressure as well as its height and depth increase with densities. Increasing the bulk density leads to a decreased local chemical potential of confined system. Therefore, the tendency of the molecules to enter the pore will increase. Consequently, the average density of the molecules inside the pore and local normal pressure will increase correspondingly. The data reported in this paper indicate when the pore width increases, the number of normal pressure oscillations increase but the heights of additional oscillations decrease.